\documentclass[12pt]{article}
\usepackage{epsfig}
\usepackage{amsmath}
\usepackage{xcolor}
\usepackage{hyperref}

\title{An alternative explanation of the 
orbital expansion of Titan and other bodies in the Solar system}
\author{Michal K\v r\'{\i}\v zek\\
{\small Institute of Mathematics, Czech Academy of Sciences, \v Zitn\'a 25,}\\
{\small CZ-115 67 Prague, Czech Republic, e-mail: krizek@math.cas.cz}\\
Vesselin G.~Gueorguiev\\
{\small Institute for Advanced Physical Studies, Sofia 1618, Bulgaria,}\\
{\small \& Ronin Institute for Independent Scholarship, 127 Haddon Pl.,}\\
{\small Montclair, NJ 07043, U.S.A., e-mail: Vesselin@MailAPS.org}\\
Andr\'e Maeder\\
{\small Geneva Observatory, University of Geneva, Chemin de Maillettes 51,}\\
{\small CH-1290 Sauverny, Switzerland, e-mail: andre.meader@unige.ch}\\
}

\begin{document}
\date{\today}
\maketitle

\abstract{Recently it was found from Cassini data that the mean 
recession speed of Titan from Saturn is $v=11.3\pm 2.0$ cm/yr 
which corresponds to a tidal quality factor of Saturn $Q\cong 100$
while the standard estimate yields $Q\ge 6\cdot 10^4$.
It was assumed that such a large speed $v$ is due to a resonance locking mechanism of 
five inner mid-sized
moons of Saturn. In this paper, we show that an essential part of 
$v$ may come from a local Hubble expansion, 
where the
Hubble-Lema\^{\i}tre constant $H_0$ recalculated to the Saturn-Titan distance $D$ is 8.15 cm/(yr\,$D$). 
Our hypothesis is based on many other observations showing a slight expansion 
of the Solar system and also of our Galaxy at a rate comparable with $H_0$.
We demonstrate that the large disproportion in 
estimating the $Q$ factor can be just caused by the local expansion effect.}

\medskip


\section{Introduction}\label{Introduction}

There is a tendency among cosmologists to state that the Hubble expansion 
can only manifest itself on cosmological scales by definition. 
Therefore, it is believed that the Universe expands globally, but not locally.
Below we show that the Universe expands at least somewhere locally. 

The Hubble expansion of local systems
has a long history dating back to the paper \cite{McVittie}. 
There are at least ten examples, even in the Solar system, 
that some tiny expansion is present.
Introducing an additional effective fictitious repulsive force,
to express the corresponding effects of local expansion,
a number of classical  paradoxes in the Solar system can be explained. 
For example, the Faint Young Sun Paradox
(first mentioned in \cite{Donn}), the very large orbital angular
momentum of Titan, Triton, Charon, and our Moon,  
rivers on Mars, 
the Tidal Catastrophe Paradox of the Moon, the slow rotation of Mercury, 
the absence of its moons, migration of planets, etc. 
-- for details see e.g. \cite{B78, A22, B95}.

The {\it Hubble-Lema\^{\i}tre constant}, which
describes the present expansion rate of the Universe, in century (cy) units is
\begin{equation}
H_0 \approx 70 \,{\rm km}/({\rm s}\,{\rm Mpc)}\approx 20\,{\rm km}/({\rm s\,Mly}) 
 \approx0.7\cdot 10^{-8} {\rm cy}^{-1}.                                                        \label{1}
\end{equation}
This is the value we will adopt in the rest of the text but for general reader we would like to point out that 
there is about 10\% variation in the value of the expansion rate depending on the type of observations utilized
to deduce the value. For example, Planck Collaboration from CMB + $\Lambda$CDM deduce that
$H_0\approx 67.4 \pm 0.5 $ km/(s\,Mpc) \cite{Planck}, 
while Gaia + HST for Cepheids and RR Lyrae yields
$H_0\approx 73.52 \pm 1.62$ km/(s\,Mpc) \cite{Riess}.

Since light travels 1 au with speed $c$ in about 500 seconds, 
we can recalibrate $H_0$ by (\ref{1})
to the astronomical unit $1\,{\rm au} = 149\,597\, 870\,700\,{\rm m}$ 
as follows
\begin{equation}
H_0 \approx \frac{0.02}{c}\,{\rm m\,s}^{-1}{\rm yr}^{-1}=
0.02\frac{500}{\rm au}{\rm\,m\,yr}^{-1}=
10\,{\rm m}\,{\rm yr}^{-1}{\rm au}^{-1}.                                                  \label{2}
\end{equation}

In Section \ref{examples}, we present several typical examples illustrating that 
this expansion rate is of the same order as some observed phenomena and 
measured data in the Solar system.
Such an expansion rate (\ref{2}) cannot be explained 
by the decrease of the solar mass \cite{Noerdlinger},
nor by the solar wind \cite[p.\,204]{A22}, 
nor by tidal forces \cite[p.\,606]{Vokrouhlicky}.
This, of course, contradicts Kepler's laws, and thus also the law of conservation of energy, 
taking into account that the Solar system is sufficiently isolated from the gravitational influence of other stars. 
For instance, the closest star (except for our Sun) Proxima Centauri acts on the Earth 
with about a million times smaller gravitational force than does Venus.

The fictitious repulsive force, which appears responsible for the expansion of the Solar system and 
other gravitationally bounded systems, is not a new fifth physical force.
It is only a side effect of gravitational forces.
There are various possible phenomena at the origin of the observed expansion, as discussed in Section \ref{TheOrigin}.
This fictitious repulsive force also influences the expansion of the Universe.  
Thus, there is no reason to assume that it would somehow avoid the Solar system.
We also give some further arguments for the existence of 
a fictitious repulsive force caused by the finite speed of gravity in Section \ref{aberration},
while in Section \ref{Cosmic considerations} we mention some cosmological models 
\cite{Capozziello, Dumin4, Iorio1, Maeder4} that could also be at the origin of local expansion. 
In Section \ref{un-proper time} we discuss the un-proper time parametrization view point.
In Section \ref{tidal quality factor}, we shall deal with a large disproportion in estimating the 
tidal quality factor $Q$ in the case of Saturn.
Finally, Section \ref{Conclusions} summarizes our conclusions.

\section{Typical examples of the local Hubble expansion 
-- observational evidence}\label{examples}

In \cite{A22}, a number of astrobiological, astronomical, 
geometrical, geochronometrical, geophysical, heliophysical, climatological, 
paleontological, and observational arguments is presented
showing that the Solar  system slowly expands by a speed of order~(\ref{2}). 
Let us briefly present some of them.


\subsection{The Earth-Moon system}

According to measurements by means of laser retroreflectors 
installed on the Moon, the average distance $d = 384\,402$ km 
between Earth and the Moon increases about 3.84 cm/yr, see \cite{Dickey}. 
This results in a total Hubble-like expansion per century (cy):
\begin{equation}
\frac{\dot{d}}{d}=\frac{~3.84\,\textrm{cm\;yr}^{-1}}
{384\,402\,\textrm{km}}=0.999\cdot 10^{-8} \textrm{cy}^{-1},                        \label{3}
\end{equation}
where the dot stands for the time derivative.
Only 55\% of the Moon's recession of 3.84 cm/yr  can be explained by tidal forces 
(see \cite{Dumin1, B61, Peltier, Flandern} for details).
The missing part $v=1.72\,{\rm cm}\,{\rm yr}^{-1}$
could be due to the local Hubble expansion or the overall assessment of the 
tidal effect could be incorrect by a factor of 2. 
From (\ref{2}) we see that this value is comparable with the average value of
the Hubble-Lema\^{\i}tre constant recalibrated to the distance 
$d=2.57\cdot 10^{-3}$ au,  namely
\begin{equation}
H_0=1000\,{\rm cm}\,{\rm yr}^{-1}{\rm au}^{-1} = 2.57\,{\rm cm}\,{\rm yr}^{-1}d^{-1}.\label{4}
\end{equation}
The corresponding expansion rate due to fictitious repulsive force is 
\begin{equation}
H_0^{\rm (loc)}=\frac{3.84(1-0.55)}{2.57}H_0= 0.67\,H_0.                              \label{5}
\end{equation}
A similar value was derived independently by 
Yurii Dumin \cite{Dumin3},
$$
H_0^{\rm (loc)} \approx 0.85\,H_0.
$$
Thus we see that the Earth-Moon system is expanding 
at significant fraction of the cosmological expansion of the Universe.

If the Earth's rotation slows down on average about 1 ms per century \cite{Maeder5},
then by the angular momentum conservation law
of the Earth-Moon system the average distance $d$ increases
about 1.25 cm/yr using the advance of \cite[pp.\,184--187]{A22}.
From this and the measured value $\dot{d}$ we get 
$3.84-1.25=2.59$ cm/yr which by (\ref{4}) yields
for the remaining part of expansion that
$$
H_0^{\rm (loc)} \approx H_0.
$$ 
This relation also follows directly from (\ref{1}) and (\ref{3}).

\smallskip  

\subsection{The Earth-Sun system}

According to Krasinsky and Brumberg \cite{Krasinski}, the present 
increase of the Earth-Sun distance is only 15 cm per year. 
However, their conclusion is derived under a nonrealistic supposition that the Newtonian 
theory of gravitation describes movements in the Solar system precisely on long-term time intervals. 
They solve an algebraic system for 62 unknown Keplerian elements of all planets 
and some large asteroids. No modeling error is assumed, no discretization error 
analysis is done, and the influence of a possible local expansion is not taken 
into account. Such small value (15 cm/yr) cannot result in relatively stable conditions 
for origin and evolution of life on Earth. The larger value (\ref{2}), 
however, could provide the needed  stable conditions for the origin and 
evolution of life on Earth, which has existed continually for at least 3.5 Gyr. 

In \cite[p.\,61]{B95}, it is proved that the average recession speed
\begin{equation}
v = 5.2\,{\rm m}\,{\rm yr}^{-1}                                                        \label{6}
\end{equation}
of the Earth from the Sun (cf.~(\ref{2}))
guarantees an almost constant solar flux on the Earth during 
the last (and also next) 3.5 billion years, since the luminosity of the Sun 
slowly and continually increases from $0.7L_0$ to the 
present value of the solar constant, see \cite{Kump},
$$
L_0 = 1.361\,{\rm kW\,m}^{-2}.
$$  
To ensure favorable conditions for life on Earth it is necessary at present that 
variations of the luminosity to be within
5\,\% of $L_0$. Note that a permanent reduction of the Sun's luminosity 
of more than 5\,\% would cause an overall glaciation of the planet.
On the other hand, at temperatures over $57\,^\circ$C
DNA sequences of many multicellular organisms decay.
The corresponding ring -- popularly called the {\it habitable zone} 
(or {\it ecosphere}) -- with radii 
$0.95^{1/2}$ au and $1.05^{1/2}$ au thus represents a very narrow 
interval 145.8--153.3 million km from the Sun.
In \cite[p.\,216]{A22},  two-sided estimates have been derived for
the recession speed $v\in[4.26,6.14]$ m/yr guaranteeing that 
the luminosity changes about 5\,\% from $L_0$, i.e., 
the Earth remains in the ecosphere during the last 3.5 Gyr.

The real mean recession speed could be even higher than (\ref{6}), 
since by data on the occurrence of fossil thermophilic bacteria
\cite{Lineweaver} the temperature of oceans was about $80\,^\circ$C about 3.5 Gyr ago.
In other words, the amount of fictitious repulsive force that is continually 
produced by the Earth-Sun system lies in a very narrow interval that 
enabled the origin of life.
Zhang, Lie, and Lei in their seminal paper \cite{Zhang} investigated 
solar and lunar data of hundreds of fossil patterns 
from the whole world. They present two other independent 
paleontological arguments 
showing that the Earth-Sun distance increases several meters per year.

\smallskip   

\subsection{The Mars-Sun system}

In 1979, Maeder and Bouvier \cite[p.\,88]{Maeder3} studied the weak field consequences of 
scale invariance of gravitation and suggested the possibility  that the recession
speed of Mars from the Sun should be several meters per year. 
Also by \cite{B78} and \cite{A22}, Mars was closer to the Sun 
from 3 to 4 Gyr ago when there were rivers on its surface.
At that time the luminosity of the Sun was about $0.75L_0$.
If Mars were to be 225 million km from 
the Sun when there was liquid water, then the corresponding 
{\it solar constant for Mars} would only be
\begin{equation}
L_{\rm Mars} = 0.75L_0\,\frac{150^2}{225^2} = \frac{L_0}{3},                     \label{7}
\end{equation}
since the solar power decreases with the square of the distance from the Sun.
The huge decrease 66.6\% of luminosity as given above thus excludes the existence 
of rivers on Mars, if it were on the same orbit as now. 
Moreover, Mars had a larger albedo than now, since there 
were water clouds feeding hundreds of large rivers (see \cite{A22}). 
It is a possibility that Mars had escaped to 
freezing in the past thanks to a strong greenhouse effect, 
which could have compensated for the lower solar luminosity (\ref{7}).
However, the absence of carbonates in lake sediments according to analysis performed by 
the Curiosity Rover in Gale Crater of Mars \cite{Bristow} is casting serious doubts on this interpretation.
For a rapid outgassing of the Martian atmosphere during the first 400 million years we refer
to \cite[p.\,16]{Scherf}. 
We further note that if Mars were to be 180 million km 
rather than 225 million km 
from the Sun when it originated, then its average recession 
speed would be 10 m/yr which is again comparable to the value of 
the Hubble-Lema\^{\i}tre constant as given in~(\ref{2}).

\smallskip  

\subsection{The Saturn system and beyond}

According to two independent astrometric and radiometric 
Cassini data \cite{Lainey1, Lainey2}, the
mean recession speed of Titan from Saturn is 
\begin{equation}
v\approx 11.3{\rm ~cm/yr}.                                                          \label{8}
\end{equation}
Before these measurements, astronomers assumed that 
this speed is only 0.1 cm/yr, because Saturn is mostly a gaseous 
planet and  thus tidal forces are small.
(Note that oceanic tides play a predominant role in the secular deceleration 
of the Earth's rotation, see \cite{Pertsev}.)
Since the Titan-Saturn average distance is 
$D = 1\,221\,870$ km, the value (\ref{8}) results in:
\begin{equation}
\frac{\dot{D}}{D}=\frac{11.3{\rm ~cm/yr}}{1\,221\,870\,{\rm km}}
=0.925\cdot 10^{-8} \textrm{cy}^{-1},                                              \label{9}
\end{equation}
by means of (\ref{2}) the recalibrated value of the Hubble-Lema\^{\i}tre constant is
\begin{equation}
H_0 = 8.15\,{\rm cm}\,{\rm yr}^{-1}D^{-1}.                                          \label{10}
\end{equation}
Then the corresponding expansion rate is
\begin{equation}
H_0^{\rm (loc)}=\frac{11.3}{8.15}H_0=1.38\,H_0.                                     \label{11}
\end{equation}
This value, of course, includes 
tides which have to be subtracted like in the case of 
our Moon, see (\ref{5}). Moreover, resonances with mid-sized
inner moons should also be taken into account.

\smallskip  

Other satellites of planets are also affected by fictitious repulsive forces. 
In particular, there exist 11 fast satellites 
that are below the stationary orbit of Uranus, where the 
orbital and rotational periods are equal. 
These satellites should approach their mother planet 
along spiral trajectories due to tidal forces. 
However, a fictitious repulsive force acts in the opposite direction 
and has approximately the same size as tidal forces. This makes 
trajectories of fast satellites stable for billions of years 
(see \cite[p.\,227]{A22} for details).
The fictitious repulsive forces, thus, prevent all fast satellites from 
crashing into their mother planets.

\smallskip 

It is an open problem how Neptune could be formed as far away as 30 au 
from the Sun, where the original protoplanetary disc was relatively
sparse and all motions are very slow (by Kepler's third law its mean speed 
is only 5.4 km/s). Assuming that the average Neptune-Sun distance increases 
roughly at the rate comparable with (\ref{2}), we find
(as in previous examples) that Neptune could be formed about 10
au closer to the Sun $4.5$ Gyr ago and
then migrate due to the fictitious repulsive force to its actual orbit. Resonances with
Jovian planets could, of course, also play an essential role in this process.
	 
\smallskip 

Let us present several more examples.
There is a well-known puzzle, how the retrograde moon Triton was captured
so far from Neptune on almost circular orbit with large radius
$a=354\,759$ km. We claim that Triton was captured closer and 
then migrated from the gaseous planet Neptune, since the fictitious repulsive 
force can be bigger than tidal forces that act in opposite direction. 
This could explain the present very large Triton's orbital angular momentum.

Similar arguments can be applied to the Pluto-Charon system.
Since this system is completely locked into the 1:1:1 resonance, 
tidal forces are negligible and its large orbital momentum can again be due to
the fictitious repulsive force.
For more information about the local Hubble expansion we refer to 
\cite{Dumin1, Dumin3, B78, A23, A26, A28, B95, Maeder1, Maeder2, Maeder3, Maeder4, Maeder5}.

In \cite{A22}, several arguments are presented showing that 
also galaxies themselves slightly expand at rate comparable with $H_0$.
For instance, by \cite[Sect.~8]{Cristina} 
the observed conservative expansion rate of the Milky Way is 0.6--1 kpc/Gyr,
which~is approximately 0.6--1 km/s, since 1 kpc $=3.086\cdot 10^{16}$ km
and 1~Gyr $=3.156\cdot 10^{16}$~s. This expansion rate nicely fits to
the Hubble constant recalculated by (\ref{1}) on the radius $R=50\,000$ ly $=15\,328$ pc
of our Galaxy, namely, $H_0=1{\rm~km/(s}\,R)$ and
\begin{equation}
\frac{\dot{R}}{R}=\frac{1{\rm ~kpc/Gyr}}{15\,328\,{\rm pc}}
=0.652\cdot 10^{-8} \textrm{cy}^{-1}.                                            \label{12}
\end{equation}
This value is again comparable with similar ratios (\ref{3}) and (\ref{9}) 
which include effects of tidal forces.
Some of the above arguments are summarized in Table \ref{Table1}. Further arguments are given in Section \ref{TheOrigin}.

\begin{table}[h]
\begin{center}
\begin{tabular}{|l|l|l|}
\noalign{\hrule}
system & $H_0^{\rm (loc)}/H_0$ & references \cr        
\noalign{\hrule}
Earth-Moon   & ~~~0.5--1$^*$     & \cite{Dumin1,Dumin3,B61} \cr
Sun-Earth    & ~~~0.5--1         & \cite{B78,A22,Zhang}     \cr
Sun-Mars     & ~~~0.3--1         & \cite{B78,Maeder3}       \cr 
Saturn-Titan & ~~~$\approx 1.3^*$& \cite{Lainey1,Lainey2}   \cr
Sun-Neptune  & ~~~$\approx 1$    & \cite{A22,B95}           \cr
Galaxy       & ~~~0.6--1$^*$     & \cite{A22,Cristina}      \cr
\noalign{\hrule}
\end{tabular}
\end{center}
\caption{\small Approximate relative values of the observed local Hubble expansion for various systems.
The values with stars are based on direct measurements/observations related to 
the Earth-Moon system, Saturn-Titan system, and our Galaxy, while the rest are indirectly deduced.}
\label{Table1}
\end{table}%

\section{The origin of the local Hubble expansion}\label{TheOrigin}

The validity of physical laws is verified by measurements. Nevertheless, 
absolutely exact measurement instruments 
cannot be constructed. Hence, we are not able to check 
by measurements whether generally accepted physical laws 
are valid for an arbitrary number of decimal digits. 
For instance, the law of conservation of energy
belongs to the basic pillars of current physics. 
However, is this law valid also in the Solar system or in our Galaxy that are only modeled by Newton's theory? 

The Newtonian gravitational law represents only a 
certain idealization of reality. It yields reliable results on short time
intervals and should not necessarily be applied on long term intervals due to large modeling errors. The 
Newtonian theory of gravitation is formulated so that the laws of conservation of energy 
and angular momentum are valid absolutely exactly. Anyway, it
does not satisfy the Principle of causality because of 
infinite speed of gravity, which causes that gravitational interaction immediately gets 
outside the corresponding light cone. 
However, the Principle of causality should 
have a priority to the laws of conservation of energy and angular momentum.

\subsection{Gravitational aberration considerations}\label{aberration}

Retardation effects in the theory of gravity were considered already in the 1898 paper 
\cite{Gerber} by Paul Gerber. Nevertheless, at present they are usually neglected in astrophysical modeling, 
see e.g. \cite{Yahalom}. Another approach which satisfies the Principle of Causality
and which does not satisfy the laws of conservation of energy and angular
momentum was first introduced by Sir Arthur Eddington in \cite[p.\,94]{Eddington}. 
To explain his main idea,
consider for simplicity only two bodies $A$ and $B$ of equal point masses $m_A=m_B>0$ that
orbit symmetrically with respect to their center of gravity.
If $A$ attracts $B$ and $B$ also
attracts $A$ at their instantaneous positions, 
then by the Newtonian theory of gravitation 
these forces are in the same line and in balance.

\begin{figure}[htbp]
\begin{center}
\includegraphics[width=.45\textwidth]{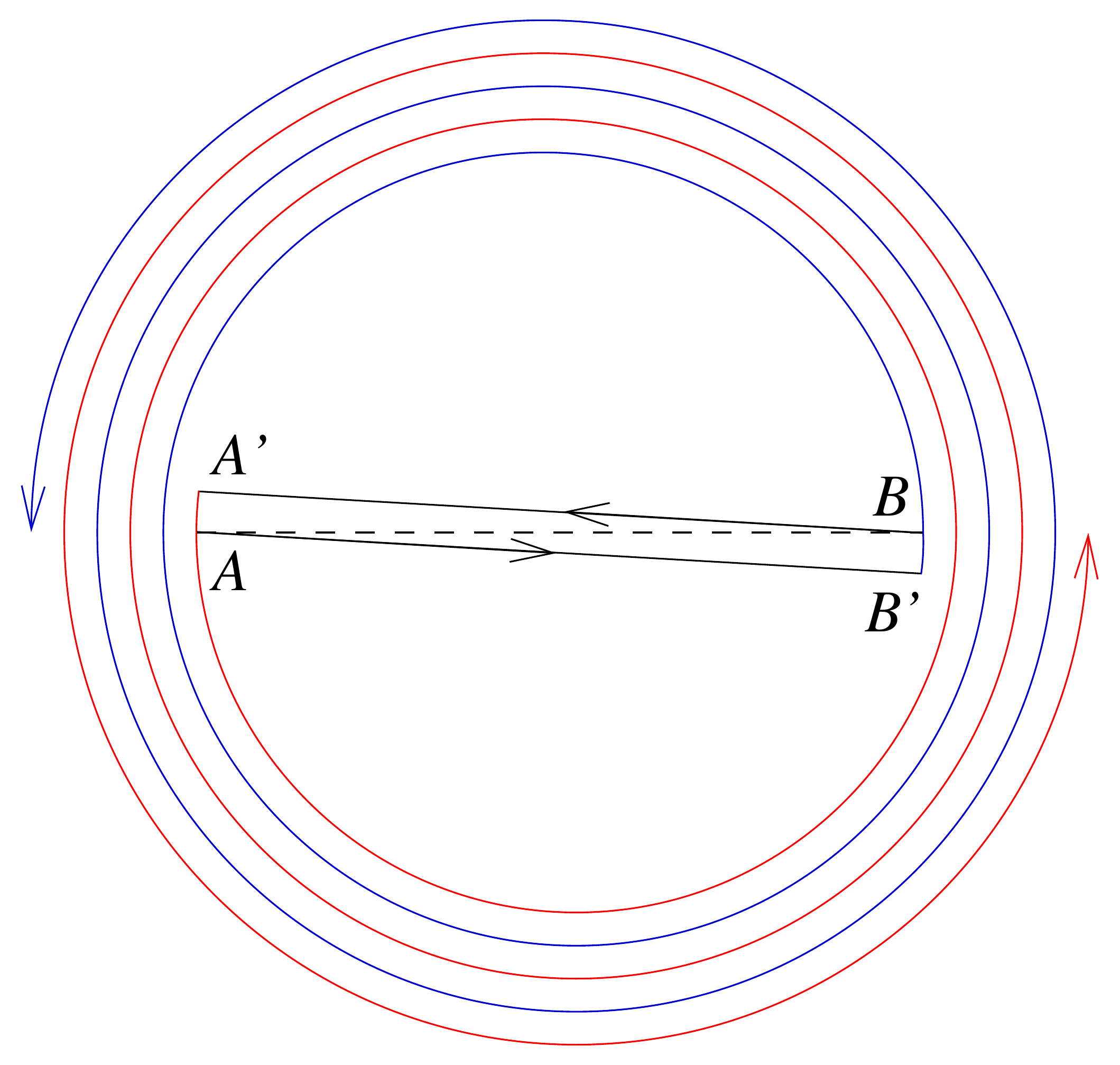}
\caption{\small Schematic illustration of trajectories of two interacting bodies 
of equal masses that form a double spiral.  The gravitational aberration angles 
$\alpha=  ABA'$ and $\alpha= BAB'$ are extremely small but positive.}
\label{Fig1}
\end{center}
\end{figure}

However, the speed of 
gravitational interaction is finite, actually $c_{\rm g}=c$. 
When the body $A$ is in $A$', it feels the attraction from $B$ (instead of $B$')
due to time delay
as depicted in Fig.~\ref{Fig1}. Similarly  
$B$' feels the attraction from $A$.
Thus, the centripetal force is stronger by the factor $1/\cos^2\alpha$,
where $\alpha=ABA$'$=BAB$' are gravitational aberration angles. 
This favors an overall contraction of the system. Nevertheless, there is a tangential
forward component of the force, since $B$ is less ``backward" than
$B$' as seen from $A$.
Then a couple of very small non-equilibrium tangential forces 
arise that permanently
act on this post-Newtonian system.
This seems to result in increase of the total angular momentum and the total
energy of the system. Hence, the corresponding trajectories
form two very slowly expanding spirals (see Fig.~\ref{Fig1}).

Emmy Noether was a leading mathematician who proved the following theorem \cite{Noether},
which completely changed the face of physics: {\it
The energy of each isolated system is conserved 
if it possesses symmetry with respect to time translations.}
Nonetheless, this symmetry is not true for spiraling trajectories
caused by fictitious repulsive force.
Similarly, another Noether's theorem states that symmetry under rotation, 
or rotational invariance, leads to the conservation of angular momentum.
This symmetry is also slightly violated in our case.

To see how non-conservative forces could arise in the above model, 
notice that the retarded gravitational force $F_{AB'}$ is at aberration angle $\alpha$.
Thus, resulting in force $F_{AB'}\times \cos\alpha\approx F_{AB} (\cos\alpha)^{-2} \cos\alpha$  
which is larger by a factor $1/\cos\alpha$ than 
the usually considered force $F_{AB}$ along the line $AB$
that goes through the center of mass of the system.
Thus, the system is not in the usual equilibrium of gravitational 
and centrifugal forces ($F_{A}=m_{A}R_{A}\omega_{A}^2$).
Furthermore, the forward pushing tangential force 
$F_{\rm t}=F_{AB'}\times \sin\alpha$ is proportional to the orbital speed
$v$, since the aberration angle $\alpha\approx\tan\alpha =v/c$. 
The presence of velocity-dependent force $F_{\rm t}\propto v$ results in a dissipative-like energy effect 
due to the non-conservative nature of such force term. This velocity-dependent
term could also induce angular momentum non-conservation, 
and as a result possible adjustment of Kepler's third law. 

The key observation here is the presence of an additional force component along the path of the object 
with a magnitude $F_{\rm t}\approx F_{AB'}\times \frac{v}{c}$. The corresponding instantaneous 
additional acceleration is then expected to be $a_{\rm t}\approx\frac{F_{AB}}{\gamma^2 m_B}\frac{v}{c}$,
where $\cos{\alpha}\approx\gamma=\sqrt{1-(v/c)^2}$. 
Note that when expanded in powers of $v/c$ the force effect is of 3rd order in $v/c$. 
If this additional acceleration is assumed to be the fictitious acceleration due to un-proper 
time parametrization, see equation (21) of
\cite{GueorgMaeder21a},
then for a two-body system, like the one shown in Fig.~\ref{Fig1}, 
we can estimate the value of the proportionality parameter $\kappa_0$ in $a_{\rm t}=\kappa_0 v$
to be $\kappa_0\approx G\frac{M}{\gamma^2 R^2c}$.
When applied to the Earth-Moon system the result is
$G\frac{M}{\gamma^2 R^2c}\approx 0.0287\, \textrm{cy}^{-1}$.
This number is too big in magnitude to have escaped detection via the observed Lunar recession 
that suggests many orders of magnitude smaller value for $\kappa_0$ based on the 
$\dot{d}/d$ value in (\ref{3}).
Therefore, such gravitational aberration at third order in $v/c$ may not be actually present 
as discussed by Carlip  \cite{Carlip}. On other hand, the gravitational radiation effect 
is much too small since it is of fifth order in $v/c\approx 3.5\cdot 10^{-6}$;
that is, $G\frac{M}{\gamma^2 R^2c} (v/c)^2 \approx 3.3\cdot 10^{-13}\,  \textrm{cy}^{-1}$.

The above discussion seems to lead the gravitational aberration idea into a dead alley. 
However, if one embraces the ideas developed below by looking at the energy 
dissipated by the force $F_{\rm t}$, then one has
$\dot{E}=F_{\rm t}v=G\frac{Mm v^2}{\gamma^2 R^2c}=726$ PW, 
while the overall energy will be approximated by 
$E\approx E_1+E_2+U\approx 1.75 \cdot 10^{29}$ J 
neglecting the Moon rotational energy (this notation is introduced below).
Therefore, for the Earth-Moon system this results in 
$\dot{E}/E\approx 1.3\cdot10^{-4} {\rm yr}^{-1}$
which per Lunar period $T=27.32$ days gives the dimensionless number 
$T\dot{E}/E\approx 9.8\cdot10^{-6}$.
These numbers are sufficiently small to evade detection so far but big enough to be 
non-negligible compared to the gravitational radiation effect as discussed in the previous paragraph.

In the  above estimates we have used that the Earth moment of inertia is 
$I_1=8.036\cdot 10^{37}$ kg\,m$^2$ (see \cite{Bursa}, also \cite[p.\,183]{A22}) and
the corresponding rotational angular frequency is $\omega_1=7.292\cdot 10^{-5}$ s$^{-1}$.
Hence, the Earth rotational momentum is 
\begin{equation}
L_1=I_1\omega_1=5.86\cdot 10^{33} {\rm~kg\,m}^2{\rm /s}                             \label{13}
\end{equation}
and its rotational energy is $E_1 \approx \tfrac12 L_1\omega_1=2.125\cdot 10^{29}$ J.
The orbital momentum of the Moon is 
\begin{equation}
L_2=m_2d^2\Omega_2=2.89\cdot 10^{34} {\rm~kg\,m}^2{\rm /s},                       \label{14}
\end{equation}
where $m_2=7.3477\cdot 10^{22}$~kg is the Moon's mass, 
$\Omega_2=2\pi/T=2.662\cdot 10^{-6}$ s$^{-1}$ is the orbital angular velocity.
The Moon orbital energy is thus $E_2 \approx \tfrac12 L_2\Omega_2=0.385\cdot 10^{29}$~J.
The corresponding gravitational potential energy is $U=-G m_1 m_2/d=-0.762\cdot10^{29}$~J,
where $m_1=5.9722\cdot10^{24}$~kg is the Earth's mass.

In the framework of Special and General Relativity Carlip derived  
that retardation effects resulting from velocity dependent field potentials  
give rise to forces that are linear (for electromagnetism) and quadratic (for gravity)
extrapolations pointing towards the instantaneous source location \cite{Carlip}. 
He showed that ``Although gravity propagates at the speed of light in general relativity, 
the expected aberration is almost exactly canceled by velocity-dependent terms in the interaction''.
Note that this remarkable result is about the force on a test particle that 
does not alter (by definition) the overall field produced by the source.
Furthermore, Carlip argues that aberration effects become significant only at powers 
consistent with radiation effects that are expected at order $v^3/c^3$
for electromagnetism and $v^5/c^5$ for gravity.
However, if one assumes that the laws of conservation of energy and angular momentum hold exactly,
then one cannot get spiraling trajectories as in Fig.~\ref{Fig1}. 

Note that  the energy dissipated by the force $F_{\rm t}$ discussed in the previous paragraph 
can also be written in terms of the gravitational potential energy 
$F_{\rm t}v=-U\times(\frac{v}{c})\frac{v}{\gamma R}$.
It can also be viewed as energy dissipated by the receding speed against the gravitational force
$F_{\rm t}v=F_{AB'}\, v^2/c=F_{AB'}\, v_r$.
This provides an estimate of the local Hubble expansion 
$H_0^{\rm (loc)}=v_r/r\approx \frac{v^2}{c\,r}=| vF_{\rm t}/U|\approx 0.03 {\rm ~cy}^{-1}$,
which seems to be missing a factor $v/c$ to be almost the same order of magnitude with (\ref{1}).
This suggests that following Carlip’s approach and computing exactly the high-order terms in $v/c$ 
beyond the first few orders can be very informative in understanding the origin 
and level of contribution of the aberration effects.

It has been discussed that gravitational aberration is associated with very small fictitious repulsive force  \cite{B61}. 
The gravitational aberration effects can be modeled by a non-autonomous 
system of differential equations with delays. However,
the simplest post-Newtonian model produces relatively quickly expanding
trajectories. The reason is that it does not take into account 
a curved space-time (nor gravitational waves).   
The proper distance between point masses $m_A$ and $m_B$ is, in fact, larger than 
the distance of their projection into the flat coordinate Euclidean space.
This mechanism contributes to the local
Hubble expansion which seems to lead to small discrepancies 
from exact conservation of energy and angular momentum within the Solar system.
This hypothesis even has the potential to substantially explain the dark energy problem
\cite{B95, Risaliti}.

\subsection{Cosmic considerations}\label{Cosmic considerations}

In order to estimate the possible magnitude of such effects we 
now discuss the series expansion of the scale factor $a(t)$ in terms of Hubble parameter $H(t)$
and deceleration parameter $q(t)$.
The Hubble parameter is defined by the well-known relation 
\begin{equation}
H=\frac{\dot{a}}{a},                                                             \label{15}
\end{equation}
where $a$ is the expansion function and the dot denotes its time 
derivative. Differentiating $H=H(t)$ with respect to time, we find that 
$$
\dot{H}=\frac{\ddot{a}}{a}-H^2=-qH^2-H^2,                                      
$$
where $q=-\ddot{a}a/\dot{a}^2$ is the {\it deceleration parameter}. The Taylor expansion 
reads (see e.g. \cite[p.\,781]{MTW}, \cite[p.\,313]{Peebles})
\begin{equation}
a(t) = a(t_0)(1+H_0\Delta t-\tfrac12 q_0 H^2_0(\Delta t)^2+\dots),              \label{16}
\end{equation}
where $t_0$ is the present time, $H_0=H(t_0)$, $\Delta t=t-t_0$, and
$q_0=q(t_0)\approx -0.6$ is the actual value of the 
deceleration parameter. 

An extremely small time derivative of $H(t)$ for the present time  
on the scale of the Solar system is derived in 
\cite{Carrera} yielding a tiny outward 
acceleration of $2\cdot 10^{-23}\,{\rm m}/{\rm s}^2$ 
at Pluto's distance of 40 au. 
Similar very small values are given in papers
\cite[p.\,62]{Cooperstock}, \cite[p.\,435]{Dicke}, and \cite[p.\,5041]{Mashhoon}
claiming that the cosmological constant $\Lambda$
has negligible influence on the Solar system.
These papers, in fact, state that the 
quadratic term in the above Taylor expansion has almost no effect on the 
accelerated expansion of the Solar system which is true, but 
surprisingly they do not take into 
account the large value of the Hubble-Lema\^{\i}tre
constant itself (see (\ref{2})) which appears at the linear term in 
(\ref{16}). For instance, if $\Delta t=1$~yr, then by (\ref{2}) we obtain
\begin{equation}
H_0\Delta t= 10~{\rm m/au}= 0.668\cdot 10^{-10}.                                \label{17}
\end{equation}
Hence, the linear term is much larger than the quadratic term in (\ref{16}),
\begin{equation}
|H_0\Delta t|\gg \frac12|q_0| H_0^2(\Delta t)^2                                \label{18}
\end{equation}
for $\Delta t$ close to 0. Consequently, the accelerated 
expansion is nonnegligible only at
cosmological distances due to (\ref{18}). However, if 
$|\Delta t|=10$ Gyr, then by (\ref{17}) we still have
$$
|H_0\Delta t| = 0.668 > 0.134 =\frac12|q_0| H_0^2(\Delta t)^2,
$$
i.e., the linear term in (\ref{16}) essentially dominates over the quadratic term 
not only on small time scales but also at time scales of the order of the age of the 
Solar system and the Milky Way galaxy.

Thus, on the whole we see that there are various physical effects through which 
some local expansion could be generated. 
The observation of such effects will need years of data and perhaps even centuries of
accumulation of such small trends to become noticeable.

\subsection{Non-conservation effects}\label{un-proper time}

In this section we discuss how
non-conservation effects could be due to the use of un-proper time parametrization 
and can be controlled by a parameter $\kappa_0$. 
The un-proper time parametrization leads to fictitious acceleration 
proportional to the velocity of a body of the form $\kappa_0 \vec{v}$ \cite{GueorgMaeder21a}.
Similar  acceleration term is also present in the week field limit of the SIV theory \cite{Maeder3,Maeder4}. 
(For another reason see also \cite{Dumin2}.)
The value of $\kappa_0$ was suggested to be related to the aberration effects earlier in the paper (see Section \ref{aberration}),
however,  the first possibility $G\frac{M}{\gamma^2 R^2c}\approx 0.0287\,\textrm{cy}^{-1}$, 
which is at the 3rd order of $v/c$, could possibly be relatively  too large, 
while the gravitational radiation effect is too small, since it is at the 5th order of $v/c$. Nevertheless, tidal effects 
and random impacts do affect the parts of the particular planetary system.
For example, in the Earth-Moon system the leading effect is 
the loss of rotational kinetic energy 
$E \propto \mathcal I \omega^2$ (see (23) of \cite{GueorgMaeder21a})
that results in  $\dot{E}/E\approx 2 \dot{\omega}/\omega=-2\dot{T}/T$, 
where $\omega$ is the Earth rotational rate $2\pi/T$.
By (\ref{13}) and (\ref{14}) the leading term in the angular momentum
(see (24) of \cite{GueorgMaeder21a}) is the Moon orbital momentum 
$L_2=27.92\cdot 10^{33}$ kg\,m$^2$/s which is a factor of $4.8$ bigger than 
Earth rotation momentum $L_1=5.86\cdot 10^{33}$ kg\,m$^2$/s.
Thus, one has $\dot{L}\approx 2L_2\dot{d}/{d}+L_1\dot{\omega}/{\omega}+\dots$.
If one insists on the usual angular momentum conservation argument then one obtains
\begin{equation}
\frac{\dot{d}}{d}\approx -\frac{L_1}{2 L_2} \frac{\dot{\omega}}{\omega}= 
-\frac{5.86}{2\cdot 27.92} \frac{\dot{\omega}}{\omega}=0.1 \frac{\dot{T}}{T}. \label{19}
\end{equation}
When using $\frac{\dot{\omega}}{\omega}=-\frac{\dot{T}}{T}$ 
with length of the day change per century $\dot{T}\approx1$ ms/cy \cite{Maeder5} 
and the standard length of the Earth sidereal day ($T=86\,164.1$ s), which is resulting in
$\dot{T}/{T}=1.157\cdot10^{-8}{\rm cy}^{-1}$, one arrives at almost an order of magnitude $\approx8.6$
times smaller value for the Moon's recession rate than the experimentally observed rate (\ref{3}):
$$
\frac{\dot{d}}{d}\approx 0.1 \frac{1\; {\rm ms/cy}}{T}=0.1 
\frac{1\; {\rm ms/cy}}{86\,164.1\;{\rm s}}=0.116 \cdot 10^{-8} \textrm{cy}^{-1}.
$$

If we apply energy conservation:
$0=\dot{E}=2E_2\dot{d}/d-U\dot{d}/d+2E_1\dot{\omega}/{\omega}+\dots$
then one would have a difficult constraint to satisfy
$\dot{d}/d\approx-2E_1/(2E_2-U)(\dot{\omega}/{\omega})$
that implies $2E_1/(2E_2-U)\approx L_1/(2 L_2)$ 
which is not satisfied for the Earth-Moon system.

If one considers non-conservation, as discussed in sub-section \ref{aberration}, 
using $L=L_1+L_2$ and $E=E_1+E_2+U$ along with the Virial theorem $<U>\,\approx-2<E_2>$, 
then one has
\begin{equation*}
\dot{E}\approx 2E_1\frac{\dot{\omega}}{\omega} + 2E_2\frac{\dot{d}}{d}-U\frac{\dot{d}}{d}+\dots
\Rightarrow 
\dot{E}\approx 2E_1\frac{\dot{\omega}}{\omega} + 4E_2\frac{\dot{d}}{d}+\dots
\end{equation*}

$$
\kappa_0=\frac{\dot{L}}{L}\approx 2\frac{L_2}{L}\frac{\dot{d}}{d}
+\frac{L_1}{L}\frac{\dot{\omega}}{\omega}+\dots
$$

\noindent
By eliminating $\kappa_0$ we obtain\footnote{We have to use $\dot{E}/E=-\kappa_0$ 
in order to agree with the loss of mechanical energy and assuming $\kappa_0>0$.}
$$2\frac{L_2}{L}\frac{\dot{d}}{d} +\frac{L_1}{L}\frac{\dot{\omega}}{\omega}
\approx
-\frac{4E_2}{E}\frac{\dot{d}}{d}-2\frac{E_1}{E}\frac{\dot{\omega}}{\omega}.
$$
Which can further be approximated using $E\approx E_1$  and $L\approx L_2$:
$$2\frac{\dot{d}}{d} +\frac{L_1}{L_2}\frac{\dot{\omega}}{\omega}
\approx
-\frac{4E_2}{E_1}\frac{\dot{d}}{d}-2\frac{\dot{\omega}}{\omega}
$$
which simplifies to
$$\frac{\dot{d}}{d}\approx -\frac{(2L_2+L_1)E_1}{2(E_1+2E_2)L_2}\frac{\dot{\omega}}{\omega}
\approx -0.8197 \frac{\dot{\omega}}{\omega}\approx 0.948 \cdot 10^{-8} {\rm cy}^{-1}.
$$
Here we have used again $\dot{T}/{T}=1.157\cdot10^{-8}{\rm cy}^{-1}$.
This value is much closer to the measured value (\ref{3}) then one based on conservation of the
angular momentum (\ref{19}).

The Solar system and our Galaxy provide us with a unique laboratory to study 
whether the laws of conservation of energy and angular momentum are valid,
see e.g. \cite{A22, angular}.
The fictitious forces cause secular migration of planets 
and their moons of order (\ref{2}). The observed slight expansion 
of the Solar system indicates that the laws of conservation of energy and angular 
momentum are slightly violated.

\section{A large disproportion in estimating the effective tidal quality factor}\label{tidal quality factor}

From Table \ref{Table2} we see that Titan and five mid-sized inner moons of Saturn have
almost circular orbits and small inclinations with respect
to Saturn's equator (see e.g. \cite{Dougherty, Fuller, Stiles}).
All these moons have synchronous rotation, i.e., there is 1:1 resonance
between each moon's rotation and its orbital period around Saturn. 
From Table \ref{Table2} we further observe that the orbital 
angular momentum of Rhea is only 1.1\,\% of Titan's momentum and the
orbital angular momenta of other inner moons are much smaller.
The radius of Titan's orbit 
is quite large with respect
to orbits of other moons so that the resonance effects 
(like 4:2:1 resonance of Io, Europa and Ganymede, see \cite{Fuller}) 
have to be small.

\begin{table}[h]
\begin{center}
\begin{tabular}{|l|c|c|c|c|c|c|}
\noalign{\hrule}
Moon &~~~ $m$ (kg) & ~~$a$ (km) & ~~~~$e$ & ~$i$ ($^\circ$) & $\overline v$ (km/s) &$L$ (kg\,m$^2$/s) \cr
\noalign{\hrule}
Mimas     &$3.7493\cdot 10^{19}$ & ~~185\,539  & 0.0196 & $1.574$ & ~~14.28 & $9.934\cdot 10^{19}$\cr
Enceladus &$1.0802\cdot 10^{20}$ & ~~237\,948  & 0.0047 & $0.009$ & ~~12.69 & $3.262\cdot 10^{20}$\cr
Tethys    &$6.1745\cdot 10^{20}$ & ~~294\,619  & 0.0001 & $1.120$ & ~~11.35 & $2.065\cdot 10^{21}$\cr
Dione     &$1.0955\cdot 10^{21}$ & ~~377\,396  & 0.0022 & $0.019$ & ~~10.03 & $4.147\cdot 10^{21}$\cr
Rhea      &$2.3065\cdot 10^{21}$ & ~~527\,108  & 0.0013 & $0.345$ &\,~~~8.48& $1.031\cdot 10^{22}$\cr
Titan     &$1.3452\cdot 10^{23}$ &1\,221\,870  & 0.0288 & $0.349$ &\,~~~5.57& $9.155\cdot 10^{23}$\cr
\noalign{\hrule}
\end{tabular}
\end{center}
\caption{\small
Basic parameters of the relevant Saturn's moons, where
$a$ is the semi-major axis,
$e$ is the eccentricity,
$i$ is the inclination to Saturn's equator,
$\overline v$ is the mean orbital speed, and
$L=ma\overline v$ is the orbital angular momentum.}
\label{Table2}
\end{table}%

\medskip

Tidal bulges raised by planets and their moons are characterized by
the effective tidal quality dimensionless factor 
$Q:=1/|2\tan\theta|\approx 1/|2\theta|$,
where $\theta$ is the tidal lag angle, see \cite{MacDonald}. For instance,
if moon's revolution period is larger than planet's rotation period,
then the lagging tide is carried ahead of the moon by the angle $\theta>0$.
On the other hand, moons below the stationary orbit produce a negative
angle $\theta<0$. For bodies on the stationary orbit, the factor $Q$ is
not well defined.

According to \cite{Goldreich}, real values of $Q$
separate sharply into two groups. Values in the range 10 to 500 correspond
to the terrestrial planets and all moons with $\theta\neq 0$
while values 
\begin{equation}
Q\ge 6\cdot 10^4                                                            \label{20}
\end{equation}
correspond to the major planets. 
The associated 
tidal lag angles fulfill $|\theta|\in(0.06^\circ,3^\circ)$ 
and $|\theta|\leq 1.7^{\prime\prime}$, respectively.

A key problem is to estimate the factor $Q$ corresponding to Saturn,
where the reciprocal value $1/Q$ is proportional to 
the thermal dissipation exerted 
by Titan and other moons in the interior of Saturn. 
According to \cite{Fuller, Lainey1, Lainey2}, one has
\begin{equation}
\frac{\dot a}{a}=\frac{3k_2}{Q}\frac{M_{\rm moon}}{M_{\rm S}}
\Big(\frac{R_{\rm S}}{a}\Big)^5 \Omega=:\frac{b}{Q},                                          \label{21}
\end{equation}
where $M_{\rm moon}$ is the mass of the moon, $R_{\rm S}=58\,232$ km and
$M_{\rm S}=5.6834\cdot 10^{26}$ kg are the mean radius and mass of Saturn, respectively,
$\Omega=2\pi/T$ 
is the moon's mean motion, $T=15.945$ days is the orbital period of Titan, and
$k_2\approx \tfrac13$ is the dimensionless Love number of degree two which 
describes the rigidity and surface deformations of Saturn under a tidal potential.

If one is to use (\ref{20}) in (\ref{21}) for Titan, then its recession speed is estimated to be
$\dot{a}<0.2$ cm/yr, which is less than $2\%$ of the observed recession (\ref{8}).
Alternatively, if one  estimates $Q$ by using the Cassini measurements (\ref{8}), Table \ref{Table2}, 
and the left-hand side of (\ref{21}) for Titan as given by (\ref{9}) -- the result is $Q=90.5$. 
However, the theory of major planets \cite[p.\,587]{Matson} predicts much larger values for $Q$ satisfying (\ref{20}).

Lainey et al.~\cite{Lainey1, Lainey2} explain
this extremely large disproportion by resonance locking of five 
inner mid-sized moons from Table \ref{Table2}. 
They consider a
complex Love number and perform a Monte-Carlo approach.
The basic idea of
tidal migration via resonances was presented earlier in \cite{Fuller}. 
There is a 2:1 resonance between Mimas
and Tethys and also between Enceladus and Dione, but no
resonance between Titan and the other moons.

We propose another explanation that could essentially help in  solving the above 
controversial issue.
If the total Titan recession (\ref{8}) is mostly due to local Hubble expansion (\ref{10}),
then the tidal effects contribute only $11.3-8.15=3.15$ cm/yr. Since 1 yr = 31\,558\,149.54 s, 
the corresponding estimate of the $Q$ factor is by (\ref{21})
$$
Q=\frac{b\,D}{3.15}\approx 325.
$$
We see that
this value is still a few orders of magnitude smaller than the expected value (\ref{20}). 
It suggests that recession due to the tidal effects and local Hubble expansion 
are not simply additive contributions to the overall recession process.
In fact, Fig.~2 of \cite{Lainey1} is effectively
a Hubble diagram showing a local expansion with a clear 
Hubble-Lema\^itre constant.
Therefore, the local Hubble expansion provides an alternative argument for explaining 
the large disproportion in estimating the effective tidal quality factor for Saturn.

\section{Conclusions}\label{Conclusions}

The main point of this work is to emphasize the overall convergence of the observed orbital expansion
of different systems such as the Earth-Moon, Titan-Saturn with the general Hubble cosmological
expansion. The past history of the Martian climate, with many evidences of liquid water connected to the relative 
absence of carbonates, is also consistent with an expansion of the Martian orbit at the same rate. Of course,
each of these systems results from  complex dynamical and astrophysical processes, however despite the differences
of these systems, the similarity of the effects and of their relative sizes is striking.  

As to the origin of the observed expansion, we have examined a number of effects and their order of
magnitude in the framework of General Relativity, and also by considering 
effects due to fictitious acceleration either aberrational or due to un-proper time parametrization. 
The problems found locally in the Solar system may also
be related to the dark components of the Universe, so that
different cosmological models  \cite{Maeder4,Capozziello} could also possibly 
be at the origin of the observed local expansion.
For instance, according to \cite{Dumin2} and \cite{Dumin4}, the local Hubble expansion arises in
a specific cosmological asymptotics of the Robertson-Walker metric at infinity
(cf.~also \cite{Iorio1}).

The fictitious forces cause secular migration of planets 
and their moons of order (\ref{2}). The observed slight expansion 
of the Solar system indicates that the laws of conservation of energy and angular 
momentum are slightly violated. 
If this force were not to act in the Solar system, 
then conditions favorable for the development of life on Earth would exist for 
only about 1 Gyr. Intelligent life would not have had enough time 
to develop due to a continual rise of temperature on Earth, see \cite{B78, B95}.

\paragraph*{Acknowledgment.} 
The authors thank Filip K\v r\'{\i}\v zek, Jaroslav Pavlousek, and Weijia Zhang for fruitful discussions. 
M.~K\v r\'{\i}\v zek was supported by the Institute of Mathematics of the Czech Academy of Sciences (RVO~67985840).
V.~Gueorguiev is extremely grateful to his wife and daughters for their understanding and family support. 
A.~Maeder expresses his deep gratitude to his wife and to D. Gachet for their continuous support.

\renewcommand\refname

\end{document}